\documentclass[pra,twocolumn,showpacs,groupedaddress,superscriptaddress,aps,10pt]{revtex4-1}
\usepackage{bm,graphicx,amsmath}
\usepackage{placeins}
\usepackage{amssymb}
\usepackage{amsmath}
\usepackage{epsfig}
\usepackage{amssymb}
\usepackage{color}
\usepackage[pdftex,colorlinks,linkcolor=blue,citecolor=blue]{hyperref}
\usepackage{subfigure}
\newcommand{\ket}[1]{\ensuremath{\left| #1 \right\rangle}}

\newcommand{\eg}{{\it{e.g.}}}

\newcommand{\1}{\ensuremath{\left|1\right\rangle}}
\newcommand{\2}{\ensuremath{\left|2\right\rangle}}
\newcommand{\3}{\ensuremath{\left|3\right\rangle}}

\begin{document}
\title{Single-site addressing of ultracold atoms beyond the diffraction limit via position dependent adiabatic passage}

\date{\today}

\author{D. Viscor}
\affiliation{Departament de F\'{\i}sica, Universitat Aut\`{o}noma de Barcelona, E-08193 Bellaterra, Spain} 
\author{J. L. Rubio}
\affiliation{Departament de F\'{\i}sica, Universitat Aut\`{o}noma de Barcelona, E-08193 Bellaterra, Spain} 
\author{G. Birkl}
\affiliation{Institut f\"ur Angewandte Physik, Technische Universit\"at Darmstadt, Schlossgartenstra{\ss}e 7, 64289 Darmstadt, Germany}
\author{J. Mompart}
\affiliation{Departament de F\'{\i}sica, Universitat Aut\`{o}noma de Barcelona, E-08193 Bellaterra, Spain} 
\author{V. Ahufinger}
\affiliation{Departament de F\'{\i}sica, Universitat Aut\`{o}noma de Barcelona, E-08193 Bellaterra, Spain} 

\begin{abstract}

We propose a single-site addressing implementation based on the sub-wavelength localization via adiabatic passage (SLAP) technique. We consider a sample of ultracold neutral atoms loaded into a two-dimensional optical lattice with one atom per site. Each atom is modeled by a three-level $\Lambda$ system in interaction with a pump and a Stokes laser pulse. Using a pump field with a node in its spatial profile, the atoms at all sites are transferred from one ground state of the system to the other via stimulated Raman adiabatic passage, except the one at the position of the node that remains in the initial ground state. This technique allows for the preparation, manipulation, and detection of atoms with a spatial resolution better than the diffraction limit, which either relaxes the requirements on the optical setup used or extends the achievable spatial resolution to lattice spacings smaller than accessible to date. 
In comparison to techniques based on coherent population trapping, SLAP gives a higher addressing resolution and has additional advantages such as robustness against parameter variations, coherence of the transfer process, and the absence of photon induced recoil. Additionally, the advantages of our proposal with respect to adiabatic spin-flip techniques are highlighted. Analytic expressions for the achievable addressing resolution and efficiency are derived and compared to numerical simulations for $^{87}$Rb atoms in state-of-the-art optical lattices.

\end{abstract}

\maketitle

\section{Introduction}
\label{sec:INTRODUCTION}

Ultracold neutral atoms in an optical lattice with single-atom and single-site resolution constitute an ideal physical system to investigate strongly correlated quantum phases \cite{Bloch'11} which, in turn, has interesting applications in quantum optics \cite{Ginsburg'07}, quantum simulation \cite{Bloch'08} and quantum information processing \cite{atoms_opt_lat,Brennen'99,Raussendorf'01}, among others. The first approaches towards single-site addressing considered the use of lattices with relatively large site separations \cite{inc_latt_space}. However, to have access to the regime of strongly correlated systems, typical lattice spacings well below $1\,\mu$m are needed since the tunneling rate has to be comparable to the on site interactions. In this case, the diffraction limit imposes strong restrictions on the addressability of individual lattice sites. To overcome this limitation different techniques have been investigated.
For instance, spatially dependent electric and magnetic fields have been used to induce position dependent energy shifts on the atom \cite{ext_fields}, allowing for site-selective addressability. Alternatively, a scanning electron microscopy system to remove atoms from individual sites with a focused electron beam \cite{SEM} has been reported. However, in this case, atoms need to be reloaded into the emptied sites after each detection event. More recently, high resolution fluorescence imaging techniques, that make use of an optical system with high numerical aperture, have been implemented to perform {\it in situ} single-atom and single-site imaging for strongly correlated systems \cite{HRFI}. 
In this context, a single-site addressing (SSA) scheme based on focused laser beams inducing position-dependent energy shifts of hyperfine states has been theoretically \cite{Zhang'06} and experimentally reported \cite{Weitenberg'11}. In the experiment, an intense addressing beam is tightly focused by means of a high resolution optical system. This beam produces spatial dependent light shifts bringing the addressed atom into resonance with a chirped microwave pulse and eventually inducing a spin-flip between two different hyperfine levels of the atom.

On the other hand, during past years, several proposals based on the interaction of spatially dependent fields, {\it e.g.}, standing waves, with three-level atoms in a $\Lambda$-type configuration have been considered, not only for single-site addressing in optical lattices but, more generally, for sub-wavelength resolution and localization \cite{Agarwal'06,swl_dark_state,Yavuz'07,SLAP}. 
In the first approaches \cite{Agarwal'06,swl_dark_state,Yavuz'07}, a spatially modulated dark state is created by means of either electromagnetically induced transparency (EIT) or coherent population trapping (CPT) \cite{ArimondoBook'96, CPT-EIT}, which allows for a tight localization of the atomic population in one of the ground states, around the position of the nodes of the spatially dependent field.
More recently, it has been shown that the resolution achieved with those CPT or EIT based techniques can be surpassed using stimulated Raman adiabatic passage (STIRAP) \cite{Bergmann'98} processes, by means of the so-called sub-wavelength localization via adiabatic passage (SLAP) technique \cite{SLAP}. The SLAP technique relies on a position-dependent STIRAP of atoms between the two ground states of a three-level $\Lambda$ atomic system, and has additional advantages compared with CPT or EIT techniques such as (i) robustness against parameter variations, (ii) coherence of the transfer process, that allows for its implementation also in Bose-Einstein condensates \cite{SLAP}, and (iii) the absence of photon induced recoil.

In this work, we apply the SLAP technique \cite{SLAP} to ultracold atoms in an optical lattice, where single-site addressing (SSA) requires one to overcome the diffraction limit. In order to address only a single site, we use here Stokes and pump pulses with Gaussian shaped spatial distributions, with the pump presenting a node centered at the lattice site that we want to address.
Assuming that all the atoms in the optical lattice are initially in the same internal ground state and applying the standard STIRAP counterintuitive temporal sequence for the light pulses \cite{Bergmann'98}, we will demonstrate that it is possible to adiabatically transfer all the atoms, except the one at the node of the pump field, to an auxiliary ground state. We will show that this process is performed with higher efficiency and yields better spatial resolution than the CPT based techniques \cite{Agarwal'06,swl_dark_state,Yavuz'07}. Also, we will demonstrate that our addressing technique requires shorter times than in the adiabatic spin-flip technique discussed in Ref.~\cite{Weitenberg'11}, and that larger addressing resolutions can be achieved using similar focusing of the addressing fields. Moreover, our technique has the additional advantage that it can be applied between two degenerated ground-state levels.

%\textcolor{blue}{***Although, the spin-flip is also achieved by a robust and adiabatic process, our SLAP based proposal has some advantages: (i) the two hyperfine levels among which the population is transferred can be degenerated, (ii) because our efficiency is higher for low intensities, we do not need an intense laser beam (revisar aquesta afirmacio), (iii) our technique works well for a focusing pump field well above the diffraction limit ($\sim\lambda$) (revisar aquesta afirmacio), and (iv) since the addressed atom do not interact with the applied fields, neither there are effects due to the spontaneous emission nor possibility of an undesired change of state of the atom.***}

The paper is organized as follows. In Sec. \ref{sec:PHYSICALMODEL} we introduce the physical system under consideration. In Sec. \ref{sec:SINGLESITEADDRESSING}, we present a protocol to achieve SSA, and we derive analytical expressions for the spatial resolution and addressing efficiency of our technique. In addition, a comparison with CPT based techniques is provided. Next, in Sec. \ref{sec:NUMERICAL SIMULATIONS}, we perform a numerical investigation of the proposed technique for a single-occupancy optical lattice loaded with $^{87}$Rb atoms by integrating the corresponding atomic density-matrix equations. Finally, in Sec. \ref{sec:CONCLUSIONS}, we summarize the results and present the conclusions.

\section{Physical model}
\label{sec:PHYSICALMODEL}

\begin{figure}
{
\includegraphics[width=0.8\columnwidth]{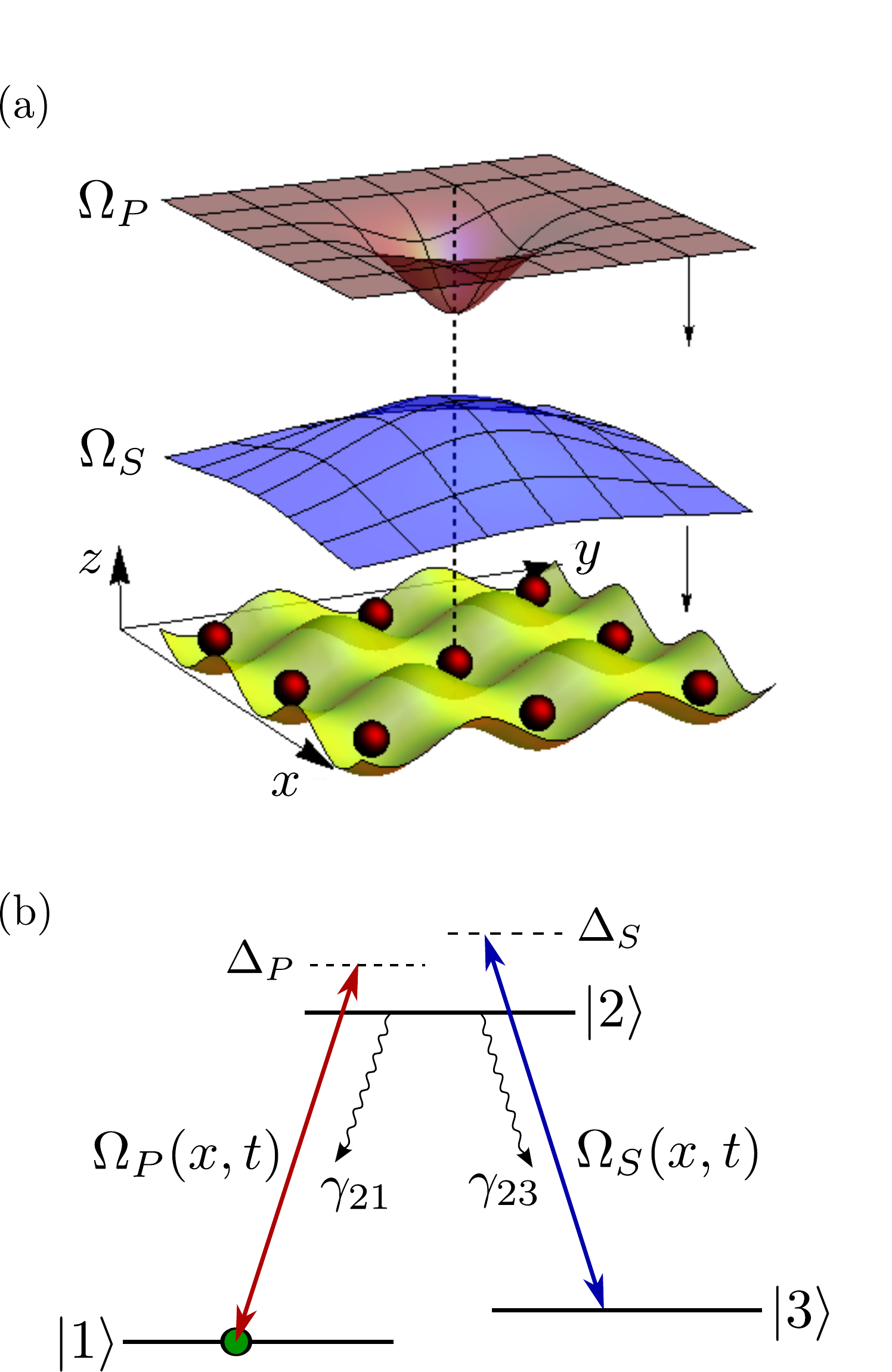}
}
\caption{
(Color online) (a) Physical system under investigation:
the pump and Stokes light pulses, with Rabi frequencies $\Omega_{P}$ and $\Omega_{S}$, propagate in the $-z$ direction and interact with the atoms of a single-occupancy optical lattice located in the ($x,y$) plane.
(b) Scheme of the $\Lambda$-type three-level atoms, initially in state \1, that interact with pump and Stokes pulses. Excited level \2 has spontaneous transition rate $\gamma_{21}$ ($\gamma_{23}$) to level \1 (\3) and $\Delta_P$ ($\Delta_S$) is the detuning of the pump (Stokes) field.}
\label{f:fig1ab}
\end{figure}

Figure~\ref{f:fig1ab}(a) illustrates the physical system under consideration. It consists of a sample of ultracold neutral atoms loaded into a two-dimensional (2D) square optical lattice with spatial period $\lambda/2$, placed in the plane ($x,y$) and illuminated by a pump and a Stokes laser pulse with Rabi frequencies $\Omega_{P}$ and $\Omega_{S}$, respectively, propagating in the $-z$ direction with a selectable time delay. 
The spatial profile of the pump pulse has a node coinciding with a particular lattice site, our target site, at which the Stokes pulse is also centered. 
With the spatial profiles of the pulses having revolution symmetry around the propagation axis [dashed line in Fig.~\ref{f:fig1ab}(a)], in the following we consider only the transverse spatial dimension $x$, without loss of generality.
In our model, we assume the system to be in the Mott insulator regime with only one atom per lattice site. Each atom is considered to have only three relevant energy levels in a $\Lambda$-type configuration, defined by the interaction with the light pulses as shown in Fig.~\ref{f:fig1ab}(b). Here, $\gamma_{21}$ ($\gamma_{23}$) is the spontaneous transition rate from the excited state $\left|2\right\rangle$ to the ground state $\left|1\right\rangle$ ($\left|3\right\rangle$) and $\Delta_P$ ($\Delta_S$) is the detuning of the pump (Stokes) field. We assume that all atoms are initially in state \1.

% % % % % % % % % % % % % % % % % % % % % % % % % % % % % % % %
% % % % % % % % % % % % % % % % % % % % % % % % % % % % %

Our approach to achieve single site addressing is based on the SLAP technique \cite{SLAP}, where, depending on their position, the atoms are transferred between two internal ground states by means of the STIRAP technique \cite{Bergmann'98}. STIRAP consists in adiabatically following one of the energy eigenstates of the $\Lambda$ system, the so-called dark state, which under the two-photon resonance condition, {\it i.e.}, $\Delta_P=\Delta_S$, has the form
\begin{eqnarray}\label{eq:darkstate}
\ket{D(x,t)}=\cos{\theta(x,t)}\ket{1}-\sin{\theta(x,t)}\ket{3},
\end{eqnarray}
where $\tan\theta(x,t)=\Omega_P(x,t)/\Omega_S(x,t)$.
Starting with all the population in \1, it is possible to coherently transfer the atomic population to state \3 changing adiabatically $\theta$ from $0^{\circ}$ to $90^{\circ}$ by means of a convenient time sequence of the fields. This time sequence corresponds to apply first the Stokes pulse and, with a certain temporal overlap, the pump pulse. Since the process involves one of the eigenstates of the system, the population transfer is robust under fluctuations of the parameter values if these are adiabatically changed and the system does not evolve near degenerate energy eigenvalues.

In the SLAP technique, the pump field has a spatial structure with nodes yielding state-selective localization at those positions where the adiabatic passage process does not occur, {\it i.e.}, those atoms placed at the nodes of the pump field remain in \1, while those interacting with both fields, pump and Stokes, are transferred to \3. For our purposes, we use the SLAP technique with a pump field having a single node at the position of the target site. Therefore, at the end of the SLAP process the population of all atoms illuminated is transferred from \1 to \3 except for the one at the node of the pump field. The spatial and temporal profiles for pump and Stokes Rabi frequencies are given by
\begin{eqnarray} 
\Omega_P(x,t)&=&\Omega_{P0}(1-\,e^{-x^2/w_P^2})\,e^{-(t-t_P)^2/2\sigma^2}, \label{eq:fieldP} \\
 %{\rm\ and\ } 
\Omega_S(x,t)&=&\Omega_{S0}\,e^{-x^2/w_S^2}\,e^{-(t-t_S)^2/2\sigma^2}, \label{eq:fieldS}
\end{eqnarray}
where $\Omega_{P0}$ and $\Omega_{S0}$ are the peak Rabi frequencies, $t_P$ and $t_S$ are the centers of the temporal Gaussian profiles, $w_P$ and $w_S$ are the spatial widths of the node in the pump and of the Stokes field, respectively, and $\sigma$ is the temporal width.

There exist several methods to create the required pump intensity profile with a central node: e.g. (i) re-imaging of a Gaussian beam with a dark central spot created by a circular absorption mask, using (ii) a Laguerre-Gaussian laser beam \cite{Heckenberg'92} or (iii) a ``bottle beam'' created by the interferometric overlap of two Gaussian beams with differing waists \cite{Isenhower'09}, or (iv) a flexible intensity pattern generated by spatial light modulators and subsequent imaging \cite{Bergamini'04,TUD_SLM_work}.

\section{Single-Site Addressing}
\label{sec:SINGLESITEADDRESSING}

In our model, we assume that the spatial wavefunctions of the individual atoms placed at the different sites, centered at $x_n$ (being $n$ the site index), correspond to the ground state of the trapping potential, which in first approximation can be considered harmonic. Therefore, the full atomic distribution in the lattice is given, initially, by
\begin{eqnarray}\label{eq:wavefunction}
\rho_{\rm \,lat}(x) = \frac{1}{w_{\rm at}\sqrt{\pi}} \sum_{n}{\exp\left[- \frac{(x-x_{n})^2}{w_{\rm at}^2} \right]},
\end{eqnarray}
where $w_{\rm at}=\sqrt{\hbar/m\omega}$ is the width of the initial atomic distribution at an individual site, $m$ is the mass of the trapped atom and $\omega$ is the harmonic trapping frequency. We assume that the addressed site is $x_{0}=0$ and their nearest neighbors $x_{\pm1}$ are at a distance $\pm \lambda/2$ where $\lambda$ is the wavelength of the fields that create the optical lattice.

In order to characterize our single-site addressing technique we consider that, once the SLAP technique has been applied, the final atomic population distribution in \1, $\rho^{\rm \,SLAP}_{1}(x)$, is given by
\begin{eqnarray}\label{eq:final distribution}
\rho^{\rm \,SLAP}_{1}(x)=P^{\rm \,SLAP}_{1\rightarrow1}(x)\rho_{\rm \,lat}(x),
\end{eqnarray}
where $P^{\rm \,SLAP}_{1\rightarrow1}(x)$ is the probability distribution that an atom remains in state \1 after the SLAP process.
Using the SLAP technique, the addressing resolution that one can obtain is related to the global adiabaticity condition \cite{Bergmann'98} at each spatial position $x$,
\begin{eqnarray}\label{eq:adiabatic condition}
\left(\Omega_{S0}e^{-x^2/w_S^2}\right)^2+\left[\Omega_{P0}\left(1-e^{-x^2/w_P^2}\right)\right]^2\geq\left(\frac{A}{T}\right)^2, \notag \\
\end{eqnarray}
where $T=t_P-t_S$ and $A$ is a dimensionless constant that, for optimal Gaussian temporal profiles and overlapping times, takes values around 10 \cite{Bergmann'98,LaA}. 
In Eq.~(\ref{eq:adiabatic condition}), the equality gives a spatial threshold $x_{\rm th}$ above which the adiabaticity condition is fulfilled. Assuming that the full width at half maximum (FWHM) of $P^{\rm \,SLAP}_{1\rightarrow1}(x)$ is $(\Delta x)_{\rm SLAP}\sim x_{\rm th}$ and expanding Eq.~(\ref{eq:adiabatic condition}) up to first order in $x$ one obtains
\begin{eqnarray}\label{eq:FWHM Prob SLAP}
\left(\Delta x\right)_{\rm SLAP}=w_S\sqrt{\frac{1+\sqrt{\left(R'+1\right)\left(\frac{A}{T\Omega_{S0}}\right)^2-R'}}{R'+1}}, \notag \\
\end{eqnarray}
where $R'\equiv R\,w_S^4/w_P^4$ and $R\equiv(\Omega_{P0}/\Omega_{S0})^2$.
Equation~(\ref{eq:FWHM Prob SLAP}) gives the width of the addressing region, and it tends to zero as $R$ increases. Moreover, since $\left(\Delta x\right)_{\rm SLAP}$ must be real valued, we find that the inequality
\begin{eqnarray}\label{eq:math limit}
\Omega_{S0} T<A\,\sqrt{\frac{1+R'}{R'}}
\end{eqnarray}
must be fulfilled. 
In this paper we will consider that $\Omega_{S0}$ is fixed, so $R'$ can be varied through $\Omega_{P0}, w_P$, and $w_S$.

Two conditions should be satisfied for our SSA technique to work. First, the population of the atom in the addressed site must remain in state \1 after the action of the fields and, second, the rest of the atoms of the lattice have to be transferred to level \3. Therefore, taking into account the overlap between $P^{\rm \,SLAP}_{1\rightarrow1}(x)$ and $\rho_{\rm lat}(x)$ in Eq.~(\ref{eq:final distribution}), it is clear that the FWHM of the probability distribution $P^{\rm \,SLAP}_{1\rightarrow1}(x)$ should satisfy 
\begin{eqnarray}\label{eq:Deltax SSA range}
(\Delta x)_{\rm at}<\left(\Delta x\right)_{\rm SLAP}<x_1-(\Delta x)_{\rm at}, 
\end{eqnarray}
where $(\Delta x)_{\rm at}=2\sqrt{\ln{2}}\,w_{\rm at}$, and $x_1=\lambda/2$ is the position of the nearest-neighboring site.
Using Eq.~(\ref{eq:FWHM Prob SLAP}), it is easy to see that these conditions fix the range for $\Omega_{S0} T$ to obtain SSA using the SLAP technique:
\begin{eqnarray}\label{eq:SSA range}
A\,\zeta_{-}<\Omega_{S0} T<A\,\zeta_{+},
\end{eqnarray}
where
\begin{eqnarray}\label{eq:limits SSA}
\zeta_{\pm}&=&\sqrt{\frac{1+R'}{\left[\left(1+R'\right)\left(\frac{x_{\pm}}{w_S}\right)^2-1\right]^2+R'}},
\end{eqnarray}
with $x_{+}=(\Delta x)_{\rm at}$ and $x_{-}=x_1-(\Delta x)_{\rm at}$. Note that the upper limit for Eq.~(\ref{eq:SSA range}) is more restrictive than Eq.~(\ref{eq:math limit}).

%%%%%%%%%%%%%%%%%
In order to have a quantitative description of the SSA performance, let us introduce the SSA efficiency as
%%%%%%%%%%%%%%%%%
\begin{eqnarray}\label{eq:TotalEfficiency}
{\eta}\equiv \mathcal{P}_{x_0}\left(1-\mathcal{P}_{x_1}\right),
\end{eqnarray}
where $\mathcal{P}_{x_0}$ corresponds to the probability of finding the atom at the addressed site $x_0$ in state \1, while $1-\mathcal{P}_{x_1}$ corresponds to the probability that the atom in the neighbor site $x_1$ has been transferred to a different internal state. We define
\begin{eqnarray}\label{eq:PartialEfficiency}
\mathcal{P}^{\rm SLAP}_{x_i}\equiv\frac{\int^{+s}_{-s}{\rho^{\rm SLAP}_1(x)dx}}{ \int^{+s}_{-s}\rho_{\rm lat}(x)dx},
\end{eqnarray}
with $\pm s=x_{i}\pm\lambda/4$ and $i=0,1$, whereas $\rho^{\rm \,SLAP}_{1}(x)$ and $\rho_{\rm lat}(x)$ have been defined in Eqs.~(\ref{eq:final distribution}) and (\ref{eq:wavefunction}), respectively.
Using Eqs.~(\ref{eq:wavefunction}) and (\ref{eq:final distribution}), the explicit forms for Eqs.~(\ref{eq:PartialEfficiency}) are
\begin{eqnarray} 
\mathcal{P}^{\rm SLAP}_{x_{0}}&=&\frac{(\Delta x)_{\rm SLAP}}{\sqrt{(\Delta x)^2_{\rm SLAP}+(\Delta x)^2_{\rm at}}}, \label{eq:ExplicitEfficiencyX0} \\
\mathcal{P}^{\rm SLAP}_{x_{1}}&=&\mathcal{P}^{\rm SLAP}_{x_{0}}\,e^{-4\ln{(2)}\,x^2_1/\left[(\Delta x)^2_{\rm SLAP}+(\Delta x)^2_{\rm at}\right]}. \label{eq:ExplicitEfficiencyX1}
\end{eqnarray} 
%
%%%%%%%%%%%%%%%%%%%%
From these expressions, it can be seen that, for $x_1>(\Delta x)_{\rm at}$, the limits given by Eq.~(\ref{eq:Deltax SSA range}), {\it i.e.}, $\left(\Delta x\right)_{\rm SLAP}=(\Delta x)_{\rm at}$ and $\left(\Delta x\right)_{\rm SLAP}=x_1-(\Delta x)_{\rm at}$, correspond to SSA efficiencies of $\sim$0.70 and $\sim$0.94, respectively.
%%%%%%%%%%%%%%%%%%%%

An alternative technique to perform atomic localization based on spatial dependent dark states is the coherent population trapping (CPT) technique \cite{Agarwal'06}. In the CPT \cite{ArimondoBook'96} technique the dark state is populated after several cycles of coherent excitation followed by spontaneous emission from \2 to the ground states. Note that, while CPT relies on spontaneous emission, the SLAP technique is fully coherent. Moreover, the latter provides higher resolution, as shown in Ref.~\cite{SLAP}, and does not suffer from recoil since the localized atoms have not interacted with light.
In what follows, we compare the range of parameters necessary to perform SSA considering both SLAP and CPT techniques. Note that we focus our comparative analysis in the CPT technique, although similar results are obtained considering the method proposed in Ref.~\cite{Yavuz'07}, where the spatial dependent dark state, \ket{D(x,t)}, is created via the STIRAP technique by switching off the fields before completing the transfer process. 

In order to compare both techniques, we define the final population distribution in \1 using CPT as $\rho^{\,\rm CPT}_{1}(x)$ in an analogous way as it has been done for the SLAP technique in Eq.~(\ref{eq:final distribution}). Then the FWHM of the corresponding probability function $P_{1\rightarrow 1}^{\,\rm CPT}(x)$ is obtained by imposing that $\left|\langle {1}\ket{D(x,t)}\right|^2=1/2$ and $t_{P}=t_{S}$ in Eqs.~(\ref{eq:fieldP}) and (\ref{eq:fieldS}):
\begin{eqnarray}\label{eq:FWHM Prob CPT} 
\left(\Delta x\right)_{\rm CPT}=\frac{2 w_{S}}{\sqrt{1+\sqrt{R'}}}.
\end{eqnarray}
Fixing the desired $\Delta x$ and using Eqs.~(\ref{eq:FWHM Prob SLAP}) and (\ref{eq:FWHM Prob CPT}) for SLAP and CPT, respectively, the constraints for the relevant parameters for each technique can be obtained. For simplicity, we consider that the Stokes pulse parameters $w_S$ and $\Omega_{S0}$ are fixed, and only the node width $w_P$ and pump peak Rabi frequency $\Omega_{P0}$ can be varied. Note that, for the SLAP case, we have to fix also $A$ and $T$.
% % % % % % % % % % % % % % % % % % % % % % % % % % % % % 
\begin{figure}
{
\includegraphics[width=1\columnwidth]{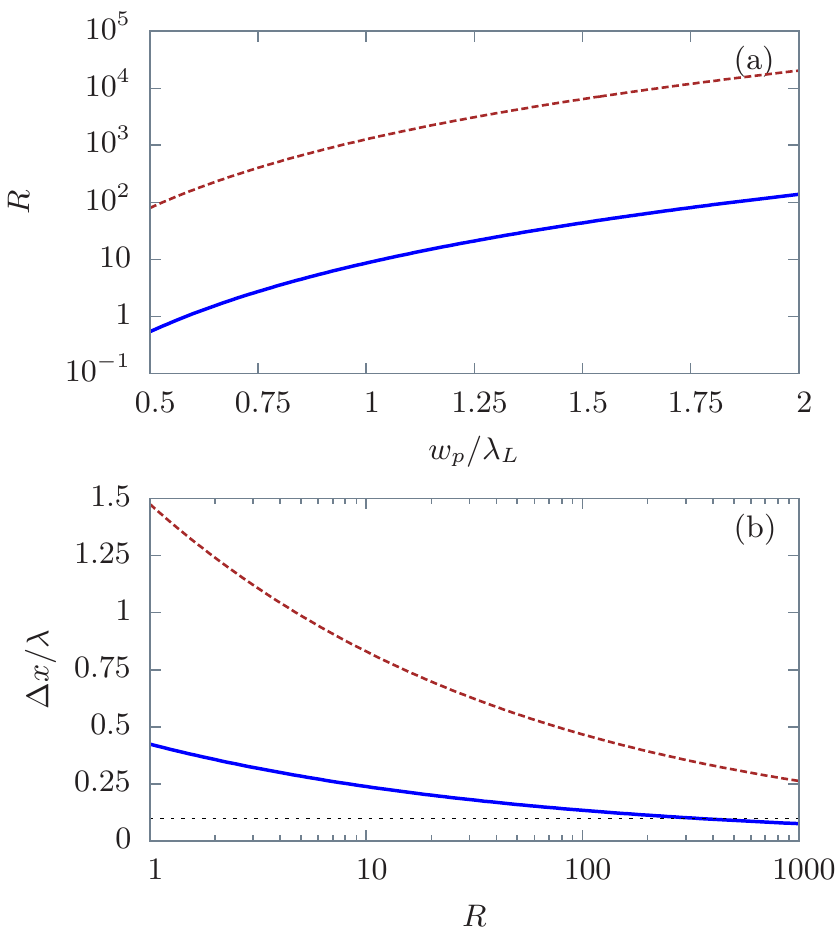}
}
\caption{
(Color online) (a) Parameter values of $R$ needed to perform SSA as a function of $w_P$, using SLAP (solid line) and CPT (dashed line) techniques. The addressing probability distribution widths are taken as $\Delta x=\lambda/4$ in both cases.
(b) $(\Delta x)_{\rm SLAP}$ (solid line) and $(\Delta x)_{\rm CPT}$ (dashed line) as a function of $R$. The FWHM of the atomic distribution, $(\Delta x)_{\rm at}=\lambda/10$ corresponds to the horizontal dotted line and we have taken $w_P=\lambda_L$. 
The parameters used in both (a) and (b) are $w_S=32\,\lambda_L=24\,\lambda$, $\Omega_{S0}T=19$, and $A=20$.
}
\label{f:fig2}
\end{figure}

% % % % % % % % % % % % % % % % % % % % % % % % % % % % %
Taking $\Delta x=\lambda/4$, half of the site separation, the required values for $R$ and $w_P$ are plotted in Fig. \ref{f:fig2}(a). To simultaneously illuminate a large number of sites, we use a large Stokes beam waist of $w_S=32\,\lambda_L=24\,\lambda$, where $\lambda_L=\frac{3}{4}\lambda$ is the wavelength of both pump and Stokes fields \cite{Note_Lambda}. The solid (dashed) line corresponds to the SLAP (CPT) case with the parameter values $\Omega_{S0}T=19$ and $A=20$ \cite{LaA}.
As $w_P$ decreases, in both SLAP and CPT cases, lower values of $R$ are needed to reach the fixed resolution $\Delta x$, since the narrower the node of the pump field, the narrower the probability distribution of atoms remaining in \1. In addition, for any given width of the node of the pump field $w_P$, the required values of $R$ are lower in the SLAP case than in the CPT case. 

It is important to realize that Eqs.~(\ref{eq:FWHM Prob SLAP}) and (\ref{eq:FWHM Prob CPT}) show the possibility to obtain, for certain parameter values, widths of the probability distribution, $(\Delta x)_{\rm SLAP}$ or $(\Delta x)_{\rm CPT}$, smaller than $(\Delta x)_{\rm at}$. In particular, using the SLAP technique this can be achieved with moderate $R$ values. This is shown in Fig.~\ref{f:fig2}(b), where $(\Delta x)_{\rm SLAP}$ (solid line) and $(\Delta x)_{\rm CPT}$ (dashed line) are represented as a function of $R$ for $w_P=\lambda_{L}$ and the rest of the parameters as in Fig.~\ref{f:fig2}(a). The FWHM of the atomic distribution, $(\Delta x)_{\rm at}=\lambda/10$, is depicted with a horizontal dotted line to indicate the values where $(\Delta x)_{\rm SLAP}<(\Delta x)_{\rm at}$. 
As we stated in the discussion of Eqs.~(\ref{eq:ExplicitEfficiencyX0}) and (\ref{eq:ExplicitEfficiencyX1}), this limit corresponds to a SSA efficiency of $\eta\sim0.70$.
This regime of parameters is interesting because it shows that the SLAP technique could be used for applications in site-selective imaging with a resolution down to the width of the atomic distribution at each site.

\section{Numerical Simulations}
\label{sec:NUMERICAL SIMULATIONS}

In this section, by numerically integrating the corresponding density-matrix equations, we study the implementation of the SLAP-based SSA technique for $\Lambda$-type three-level $^{87}$Rb atoms in a single-occupancy optical lattice. Numerical calculations using the CPT technique are also presented for comparison. 
The wavelength of the lasers that create the optical lattice, red detuned with respect to the $D1$ line of $^{87}$Rb, is $\lambda=1064$ nm. The potential depth of the optical lattice is chosen as $V_0=15E_r$, where $E_r$ is the recoil energy. This corresponds to a harmonic trapping frequency of $\omega=2\pi\times15.92$ kHz \cite{Bloch'08}. 
Therefore, the FWHM of the atom distribution at each site due to the confining potential is $(\Delta x)_{\rm at}=142$ nm.
Pump and Stokes fields with $\lambda_{L}=795$ nm are coupled to $\1\leftrightarrow\2$ and $\3\leftrightarrow\2$, respectively, where $\1\equiv\ket{F=2,m_F=-2}$, $\2\equiv\ket{F'=2,m_F=-1}$ and $\3\equiv\ket{F=2,m_F=0}$ are hyperfine energy levels of the $D1$ line of $^{87}$Rb. The excited state \2 has a spontaneous transition rate $\gamma_{21}=2\pi\times0.96$ MHz ($\gamma_{23}=2\pi\times1.44$ MHz) to state \1 (\3), and we assume no spin decoherence during the interaction time.
We consider equal temporal pulse widths of $\sigma=0.2\,\mu$s with a temporal delay $T=1.4\sigma$, in such a way that the total SSA process time is $4\sigma$. The Stokes pulse has a maximum Rabi frequency $\Omega_{S0}=19/T=2\pi\times10.8$ MHz, while the maximum Rabi frequency of the pump is varied through the parameter $R$, since $\Omega_{P0}=\Omega_{S0}\sqrt{R}$. 
Concerning the spatial profiles of the fields, we assume a wide Stokes profile, $w_S=32\lambda_L$, and a narrow node for the pump, $w_P=\lambda_L$.
%%%%%%%%%%%%%%%%%%%%%
\begin{figure}
{
\includegraphics[width=1\linewidth]{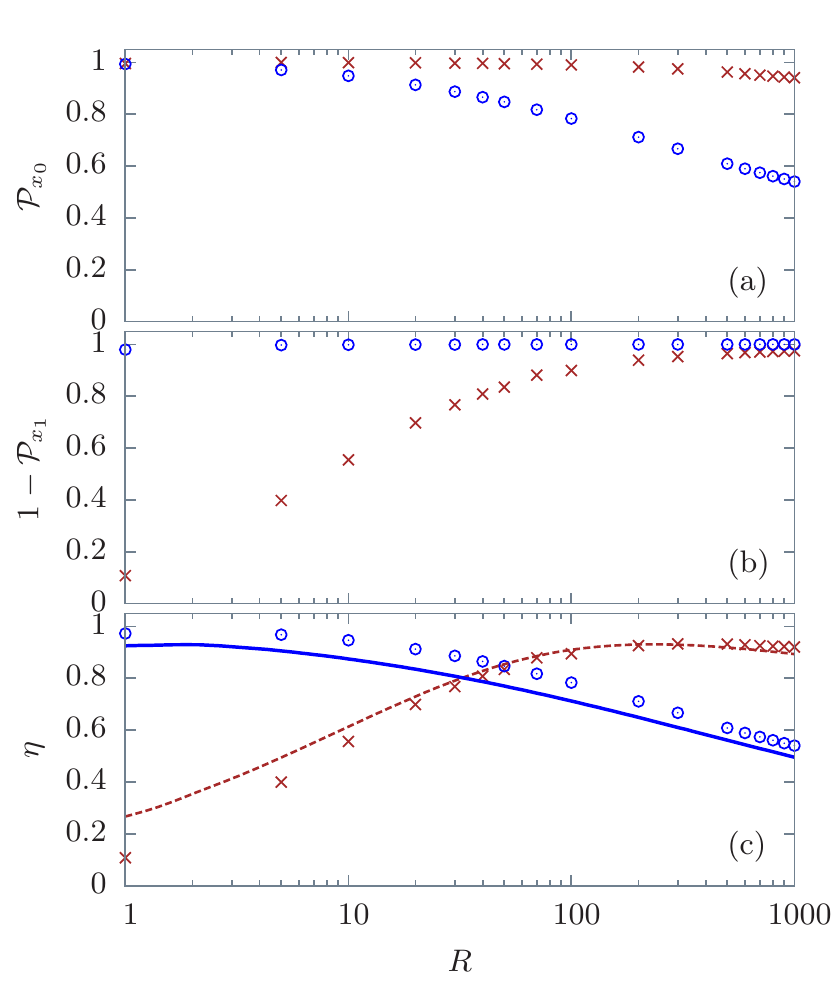}
}
\caption
{(Color online) Numerical results for the probability of finding the atom located at $x_0$ in state \1 (a), the probability to transfer it from \1 to another state (b), and the efficiency $\eta$ as a function of $R$ (c), for the SSA with SLAP (circles) and CPT (crosses) techniques. Analytical curves for SLAP (solid line) and CPT (dashed lines), computed from Eqs.~(\ref{eq:TotalEfficiency}) and (\ref{eq:PartialEfficiency}), are added in (c) for comparison (see text for the rest of the parameters).}
\label{f:EfficienciesNum}
\end{figure}
%%%%%%%%%%%%%%%%%%%%%%%%%%%%%%%%%%%%%%%%%%%%%%%%%%%%%%%%%%%%%%%%%%%%%%
As it has been discussed in the previous section, to properly perform SSA, the population of all the atoms in the lattice, except the one in the addressed site, must be transferred from \1 to \3 with high probability. 
In what follows, those requirements for the realization of the SSA are studied by numerically evaluating the SSA efficiency.

The signatures of SSA are shown in Fig.~\ref{f:EfficienciesNum}, where the numerically evaluated efficiency and probabilities defined in Eqs.~(\ref{eq:TotalEfficiency}) and (\ref{eq:PartialEfficiency}) are plotted as a function of $R$ for both the SLAP (circles) and CPT (crosses) techniques. 
Figures~\ref{f:EfficienciesNum}(a) and \ref{f:EfficienciesNum}(b) show the probabilities $\mathcal{P}_{x_{0}}$ and $1-\mathcal{P}_{x_{1}}$, respectively. For large values of $R$ the probability of finding the atom at the addressed site in \1 is higher with the CPT than with the SLAP technique [see Fig.~\ref{f:EfficienciesNum}(a)], while for small values of $R$ the probability of removing the atom from the neighboring site [see Fig.~\ref{f:EfficienciesNum}(b)] is higher in the SLAP case. 
This is explained by the fact that $(\Delta x)_{\rm SLAP}<(\Delta x)_{\rm CPT}$ for a given value of $R$, as shown in Fig.~\ref{f:fig2}(b). 
The SSA efficiency results obtained by the numerical evaluation of Eqs.~(\ref{eq:TotalEfficiency}) and (\ref{eq:PartialEfficiency}) are shown in Fig.~\ref{f:EfficienciesNum}(c), together with the corresponding analytical curves (solid and dashed lines) obtained using Eqs.~(\ref{eq:ExplicitEfficiencyX0}) and (\ref{eq:ExplicitEfficiencyX1}), added for comparison. 
A good agreement is found between numerical and analytical results. 
From Fig.~\ref{f:EfficienciesNum}(c) it is clear that the SLAP technique is more efficient for lower values of the intensity ratio of the addressing fields ($R<50$) than the CPT technique. Certainly, this is advantageous for the experimental implementation with limited laser power available.

%%%%%%%%%%%%%%%%%%%%%%%%%%%%%%%%%%%%%%%%%%%%%%%%%%%%%%%%%%%%%%%%%%%%%%
\begin{figure}
{
\includegraphics[width=1\linewidth]{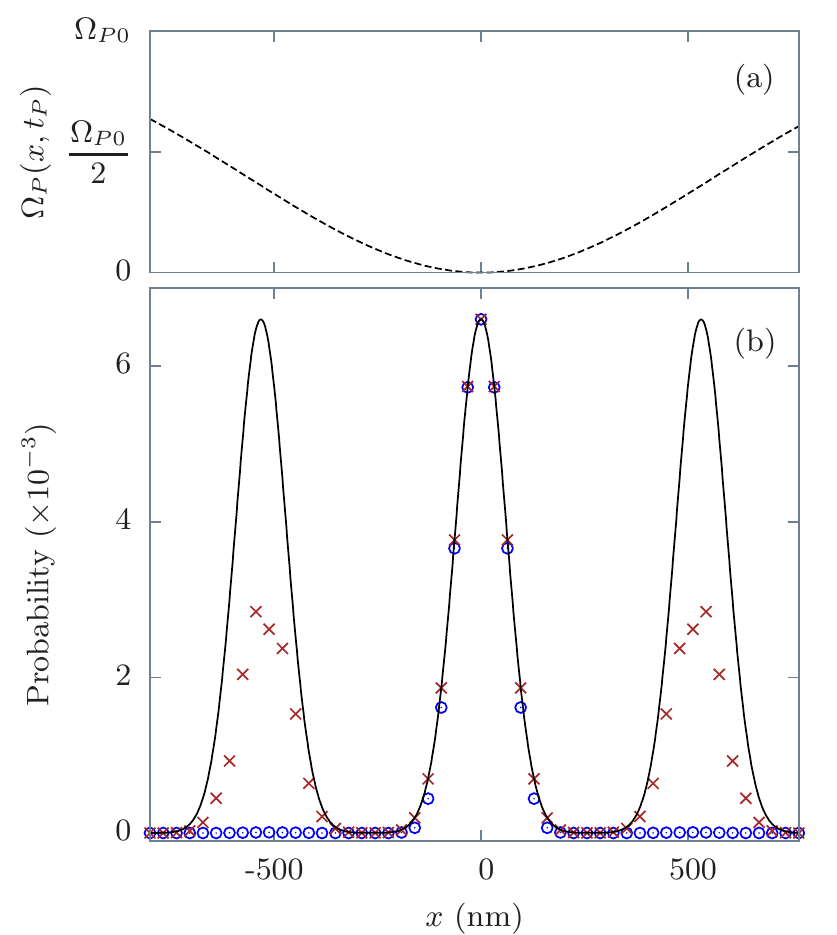}
}
\caption
{(Color online) (a) Spatial profile of the pump pulse with its node focused at $x=0$. (b) Final population distribution remaining in \1 using SLAP (circles) and CPT (crosses) single-site addressing techniques for $R=10$. The initial atomic distribution in \1, $\rho_{\rm lat}(x)$, is shown as a solid line (see text for the parameters values).
}
\label{f:RubidiumPeaks}
\end{figure}
%%%%%%%%%%%%%%%%%%%%%%%%%%%%%%%%%%%%%%%%%%%%%%%%%%%%%%%%%%%%%%%%%%%%%%

An example of the final population distribution after performing SSA with the SLAP (circles) and the CPT (crosses) techniques is plotted in Fig.~\ref{f:RubidiumPeaks}(b) for the particular case of $R=10$ and with the rest of the parameters as in Fig.~\ref{f:EfficienciesNum}. The initial population distribution in \1, $\rho_{\rm lat}(x)$, at the addressed site ($x_0=0$), and the two next neighbors $x_{\pm1}=\pm532$ nm is shown as a solid line. 
Note that, in the SLAP case (circles), the population of state \1 around $x_0$ remains almost the same after the addressing process, while in the first neighbor sites it is practically zero. On the other hand, for the CPT case (crosses), the population in the addressed site remains also nearly unchanged, but it exhibits a significant amount of population in the neighbor sites. 
This is in full agreement with the discussion following Figs.~\ref{f:EfficienciesNum}(a) and \ref{f:EfficienciesNum}(b).
For this example, the total efficiencies found are $\eta^{\rm CPT}=0.56$ and $\eta^{\rm SLAP}=0.95$ according to the corresponding values shown in Fig.~\ref{f:EfficienciesNum}(c). In addition, as shown in Fig.~\ref{f:RubidiumPeaks}(a), the width of the pump node required to perform the SSA method 
is much larger than the addressed region $(\Delta x)_{\rm SLAP}$. In particular, for the case shown in Fig.~\ref{f:RubidiumPeaks}, $w_{P}=\lambda_{L}$ and $(\Delta x)_{\rm SLAP}\simeq\lambda_L/3$, thus obtaining addressing resolution beyond the diffraction limit.

Finally, we carry out a comparison between our proposal and the experiment reported in Ref.~\cite{Weitenberg'11}, where a focused laser beam induces position-dependent light shifts, allowing one to perform a spin flip by means of a resonant microwave pulse at the addressed site.
Since the microwave field involved in the experiment has a Rabi frequency of kHz, the total spin-flip time is in the order of ms. In contrast, as our proposal makes use of only optical fields, the addressing time is three orders of magnitude below ($\mu$s). Specifically, to achieve similar values of the addressing resolution in both techniques, $(\Delta x)\sim300$ nm, we have obtained an addressing time of $\sim$40$\,\mu$s.
This decrease of the total addressing time needed implies a reduction of the effects caused by spontaneous scattering of photons, which are a limitation for the light shifts-based proposals \cite{Zhang'06, Weitenberg'11}. These effects could be strongly reduced in our case.
Also, we have compared the resolution obtained in both techniques. In Ref.~\cite{Weitenberg'11}, they use an addressing beam with an intensity FWHM of approximately 600 nm, and obtain a spin-flip probability distribution with FWHM=330 nm. 
In our technique, using $R = 1$ and a width of the pump node $w_{P}=509$ nm, which corresponds to the width of their addressing beam, a very similar FWHM of the addressing probability distribution is obtained: $(\Delta x)_{\rm SLAP}=330.66$ nm. Note that this value can be reduced by increasing the ratio between the pump and the Stokes intensities, \eg, $R = 10$ implies $(\Delta x)_{\rm SLAP}=181.86$ nm and for $R = 100$ we obtain $(\Delta x)_{\rm SLAP}=100.82$ nm.

\section{Conclusions}
\label{sec:CONCLUSIONS}

In this work we have discussed a proposal to perform single-site addressing (SSA) in an optical lattice by using the SLAP technique. 
With respect to other dark state based techniques such as CPT, this method is fully coherent, robust against variations of the parameter values, and we have found that it yields higher efficiencies for smaller values of the intensity ratio between the pump and the Stokes fields. Moreover, the addressed atom does not interact with the fields, minimizing all possible decoherent processes.
On the other hand, with respect to the recent experiment on SSA using adiabatic spin flips \cite{Weitenberg'11}, the present proposal allows one to use two degenerate ground levels, takes shorter times to perform the addressing process, and provides similar or even larger addressing resolutions for similar focusing of the addressing fields.

The proposed method provides an achievable addressing resolution that can be pushed well below the diffraction limit of the addressing light field and of the optical setup used for addressing or detection of atoms at closely spaced lattice sites.
This relaxes the requirements on the optical setup or extends the achievable spatial resolution to lattice spacings smaller than accessible to date. 
Through analytical considerations, we have derived the range of parameters for which SSA is properly achieved. Moreover, we have obtained expressions to estimate the resolution and the efficiency of the SLAP-based addressing method, and we have compared them with the analogous expressions obtained using the CPT technique. Next, by integrating the density-matrix equations with realistic parameter values for state-of-the-art optical lattices loaded with $^{87}$Rb atoms, we have checked the validity of the analytical approach.

\begin{acknowledgments}

We thank Albert Benseny for fruitful discussions and comments. We acknowledge financial support from the Spanish Ministry of Science and Innovation under Contracts No. FIS2008-02425, No. FIS2011-23719, and No. CSD2006-00019 (Consolider project Quantum Optical Information Technologies), from the Catalan Government under Contract No. SGR2009-00347, from the German Research Foundation DFG (Contract No. BI 647/3-1), and from the German Academic Exchange Service DAAD (Contract No. 0804149).

\end{acknowledgments}


\begin{thebibliography}{99}
\bibitem{LaA} Although usually the value $A\sim10$ is accepted, we have found that $A=20$ gives a more accurate description for the values of $\sigma$ and $R$ considered along this paper.
\bibitem{Note_Lambda} We choose this ratio $\lambda_L=\frac{3}{4}\lambda$ in order to be consistent with the specific numerical values used in Sec. \ref{sec:NUMERICAL SIMULATIONS}.
\end{thebibliography}
\end{document}